\documentclass[prb,twocolumn,superscriptaddress,preprintnumbers,amsmath,amssymb,aps,floatfix]{revtex4-2}

\usepackage{graphicx}
\usepackage{dcolumn}
\usepackage{bm}
\usepackage{hyperref}
\usepackage{tabularx}
\usepackage{orcidlink}
\usepackage{adjustbox}
\usepackage{multirow}

\begin{document}

\title{Oxygen sublattice disorder and valence state modulation in infinite-layer nickelate superlattices}

\author{R.~A.~Ortiz \orcidlink{0000-0001-7812-0123}}
\email{Roberto.Ortiz@web.de}
\affiliation{Max Planck Institute for Solid State Research, Heisenbergstra{\ss}e 1, 70569 Stuttgart, Germany}

\author{N.~Enderlein \orcidlink{0000-0001-9832-1109}}
\affiliation{Department of Physics, Friedrich-Alexander-Universit{\"a}t Erlangen-N{\"u}rnberg (FAU), 91058 Erlangen, Germany}

\author{K.~F{\"u}rsich \orcidlink{0000-0003-1937-6369}}
\affiliation{Max Planck Institute for Solid State Research, Heisenbergstra{\ss}e 1, 70569 Stuttgart, Germany}

\author{R.~Pons}
\affiliation{Max Planck Institute for Solid State Research, Heisenbergstra{\ss}e 1, 70569 Stuttgart, Germany}

\author{P.~Radhakrishnan \orcidlink{0000-0002-8605-3474}}
\affiliation{Max Planck Institute for Solid State Research, Heisenbergstra{\ss}e 1, 70569 Stuttgart, Germany}
\affiliation{Center for Quantum Phenomena, Department of Physics, New York University, New York, New York, 10003, USA}

\author{E.~Schierle}
\affiliation{Helmholtz-Zentrum Berlin f\"{u}r Materialien und Energie, Albert-Einstein-Stra{\ss}e 15, 12489 Berlin, Germany}

\author{P.~Wochner}
\affiliation{Max Planck Institute for Solid State Research, Heisenbergstra{\ss}e 1, 70569 Stuttgart, Germany}

\author{G.~Logvenov}
\affiliation{Max Planck Institute for Solid State Research, Heisenbergstra{\ss}e 1, 70569 Stuttgart, Germany}

\author{G.~Cristiani}
\affiliation{Max Planck Institute for Solid State Research, Heisenbergstra{\ss}e 1, 70569 Stuttgart, Germany}

\author{P.~Hansmann \orcidlink{0000-0002-0330-7927}}
\affiliation{Department of Physics, Friedrich-Alexander-Universit{\"a}t Erlangen-N{\"u}rnberg (FAU), 91058 Erlangen, Germany}

\author{B.~Keimer \orcidlink{0000-0001-5220-9023}}
\affiliation{Max Planck Institute for Solid State Research, Heisenbergstra{\ss}e 1, 70569 Stuttgart, Germany}

\author{E.~Benckiser \orcidlink{0000-0002-7638-2282}}
\email{E.Benckiser@fkf.mpg.de}
\affiliation{Max Planck Institute for Solid State Research, Heisenbergstra{\ss}e 1, 70569 Stuttgart, Germany}

\date{\today}

\begin{abstract}
The family of infinite-layer nickelates promises important insights into the mechanism of unconventional superconductivity. Since superconductivity has so far only been observed in epitaxial thin films, heteroepitaxy with the substrate or a capping layer possibly plays an important role. Here, we use soft x-ray spectroscopy to investigate superlattices as a potential approach for a targeted material design of high-temperature superconductors. We observe modulations in valence state and oxygen coordination in topotactically reduced artificial superlattices with repeating interfaces between nickelate layers and layers of materials commonly used as substrates and capping layers. Our results show that depending on the interlayer material metallic conductivity akin to the parent infinite-layer compounds is achieved. Depth-resolved electronic structure measured by resonant x-ray reflectivity reveals a reconstructed ligand field and valence state at the interface, which is confined to one or two unit cells. The central layers show characteristics of monovalent nickel, but linear dichroism analysis reveals considerable disorder in the oxygen removal sites. We observe a quantitative correlation of this disorder with the interlayer material that is important for future modeling and design strategies.
\end{abstract}

\maketitle

\section{Introduction}

The discovery of superconductivity in infinite-layer nickelate epitaxial films \cite{Li2019} has stimulated extensive new research in recent years. Much of this interest stems from the close structural and electronic similarities between nickelates and high-temperature superconducting cuprates, as the undoped infinite-layer nickelates show planes with formally Ni$^{1+}$ ions with a $3d^9$ electron configuration in a square-planar oxygen ligand field. Despite the intensive ongoing research, the synthesis and characterization of superconducting samples continue to present a significant challenge \cite{Gutierrez2024,Xu2024}. Nonetheless, several material compositions of epitaxial films with alkaline-earth doped infinite-layer structure $R_{1-x}A_x$NiO$_2$ ($R$: rare-earth ion, $A$: Ca or Sr) \cite{Osada20201,Osada2021,Zeng2022} were reported to show superconductivity below 15-20~K with a dome-like doping dependence of the superconducting phase \cite{Zeng2020, Li2020}. Significant antiferromagnetic exchange interactions \cite{Lu2021,Fowlie2022} as well as a possible charge order in the underdoped region of the phase diagram \cite{Tam2022, Krieger2022, Rossi2022} suggest similarities to the cuprates, although the latter is highly controversial \cite{Parzyck2024b}.

In addition, the influence of epitaxy with the substrate and a capping layer, which is often used for phase stabilization, is not fully understood. So far, superconductivity in infinite-layer nickelates has only been reported in epitaxial heterostructures grown on different substrates that induce different strains. Recently, it has also been possible to synthesize free-standing superconducting membranes in which the superconducting transition temperature decreases when released from a compressive strain-inducing substrate \cite{Lee2025,Yan2024}. Contradictory results have been reported for the occurrence of charge ordering depending on the presence of a capping layer \cite{Tam2022, Krieger2022, Rossi2022, pelliciari2023, Parzyck2024b, Benckiser2022}. In a recent publication, superconductivity was observed in optimally Sr-doped nickelates sandwiched between SrTiO$_3$ (STO) in superlattices above a critical thickness of the nickelate layer stacks \cite{Wen2024}. However, the authors found no evidence that the interfaces to SrTiO$_3$ had an influence and identified the superconductivity as an intrinsic property of the Nd$_{0.8}$Sr$_{0.2}$NiO$_2$ layer stacks.

In the present study, we investigate the influence of the interfaces of the parent compound infinite-layer nickelates sandwiched between layers of a second, chemically inert perovskite material in ($R$NiO$x$)$_m$-(ABO$_3$)$_n$ superlattices. Previous work has shown that such superlattices show greater structural stability and reversibility in the reduction process and in principle allow for doping without cation disorder \cite{Ortiz2021}. Assuming that the averaged nickel valence state Ni$^{1+\frac{2}{m}}$ is controlled by the number $m$ of nickelate layers, a variation of doping is possible through changing the thickness. In our previous study of (LaNiO$_x$)$_8$-(LaGaO$_3$)$_4$ superlattices, nominally with Ni$^{1.25+}$, results from x-ray absorption and density functional theory + dynamical mean-field theory (DFT+DMFT) suggested that self-doped holes originating from the interfaces with LaGaO$_3$ (LGO) remain in the interfacial nickel oxide atomic layers. This effect is theoretically explained by the formation of Hund's-coupled Ni$^{2+}$ with $S=1$ at the interface, which prevents a homogeneous hole doping of the infinite-layer stacks.

Here we extended our previous work by investigating the influence of the composition of a different interlayer material, STO and use x-ray resonant reflectometry (XRR) to obtain layer-resolved information about the local electronic structure of nickel, including its polarization dependence in a non-destructive manner. The choice of STO as interlayer material is intriguing for several reasons. Firstly, some studies reported that a STO capping layer significantly aids in phase stabilization of infinite-layer thin films \cite{Li2019,Lee2020,Osada2021,Krieger2022,Krieger2023,Parzyck2024c}. Secondly, scanning transmission electron microscopy (STEM) studies indicated that the phase-pure perovskite structure seems to stabilize more easily within a few layers near the STO-substrate interface \cite{Li2019,Lee2020}. Thirdly, the polar mismatch at the STO-nickelate interface is expected to lead to structural \cite{Bernardini2022} and/or electronic reconstructions \cite{Geisler2021}. A recent STEM-EELS study demonstrated that both p-type and n-type interfaces with STO are present in (Nd,Sr)NiO$_2$ thin films, although the influence of the reconstructions is limited to 2-3 unit cells at the interface \cite{Raji2024b}. In the superlattices that are the focus of this study, the $R$NO-STO ($R$= La, Nd) interfaces are repeated several times, thus amplifying their potential effect.

\section{Details of experiment and calculation}

\begin{figure*}[tb]
\center\includegraphics[width=0.99\textwidth]{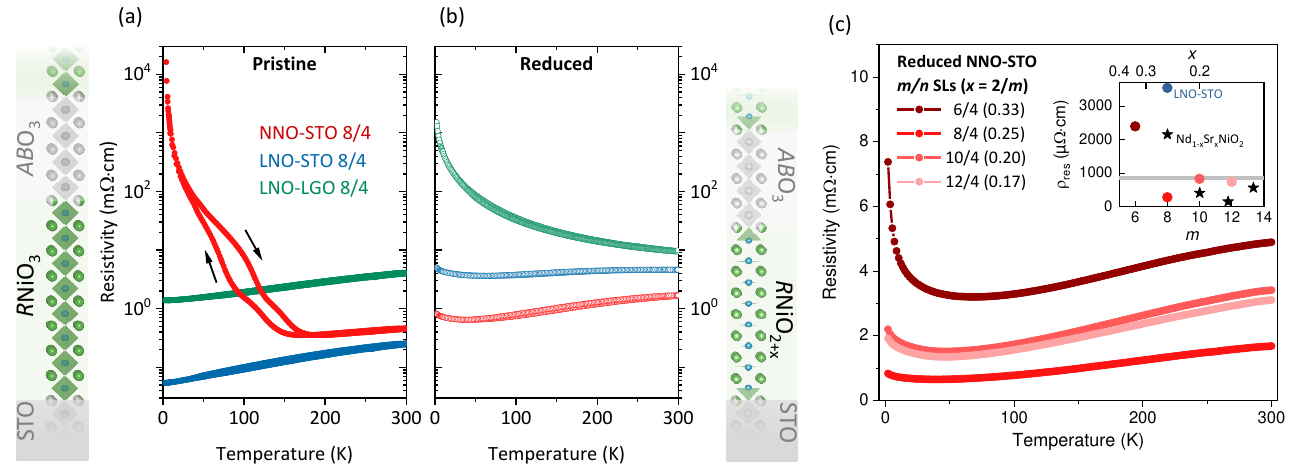}
\caption{Temperature-dependent resistivity of three pairs of (a) pristine and (b) reduced superlattices, with different compositions: LNO-STO, NNO-STO, and LNO-LGO \cite{Ortiz2021}. The measurements were performed between 2 and 300\,K. For pristine NNO-STO (red curve in (a)) a hysteresis is observed for heating (arrow pointing down) and cooling (arrow pointing up) across the metal-insulator transition in NdNiO$_3$. (c) Thickness-dependence of the resistivity upon variation of the NNO stack thickness from $m=6$ to 12 u.c. The inset displays the residual resistivity ($\rho_{\rm res}$) together with reference values that were measured on Nd$_{1-x}$Sr$_{x}$NiO$_2$ thin films with varying $x$ (upper axis) in Ref.~\onlinecite{Li2020}. The horizontal gray line indicated the quantum sheet resistance per NiO$_2$ two-dimensional plane.}\label{Fig_Resistivity}
\end{figure*}

Three superlattices with composition NdNiO$_3$-SrTiO$_3$ (NNO-STO), LaNiO$_3$-SrTiO$_3$ (LNO-STO), and LaNiO$_3$-LaGaO$_3$ (LNO-LGO) with ($m=6-12$ u.c./$n=$4 u.c.)$\times$8 stacking (u.c.= perovskite unit cells) were grown by pulsed-laser deposition onto (001)-SrTiO$_3$ substrates under the same conditions reported in Refs.~\onlinecite{Wu2013,Ortiz2021}. After growth, samples with lateral dimensions $5\times 5$\,mm$^2$ were cut into four equally sized pieces, one from which was kept pristine, and the remaining pieces were reduced. The reduction process was performed following a similar preparation procedure as described in Ref.~\onlinecite{Ortiz2021}, but with a different time range, specifically 96~h for LNO-STO and 201~h for NNO-STO.

\begin{figure*}[bt]
\center\includegraphics[width=0.99\linewidth]{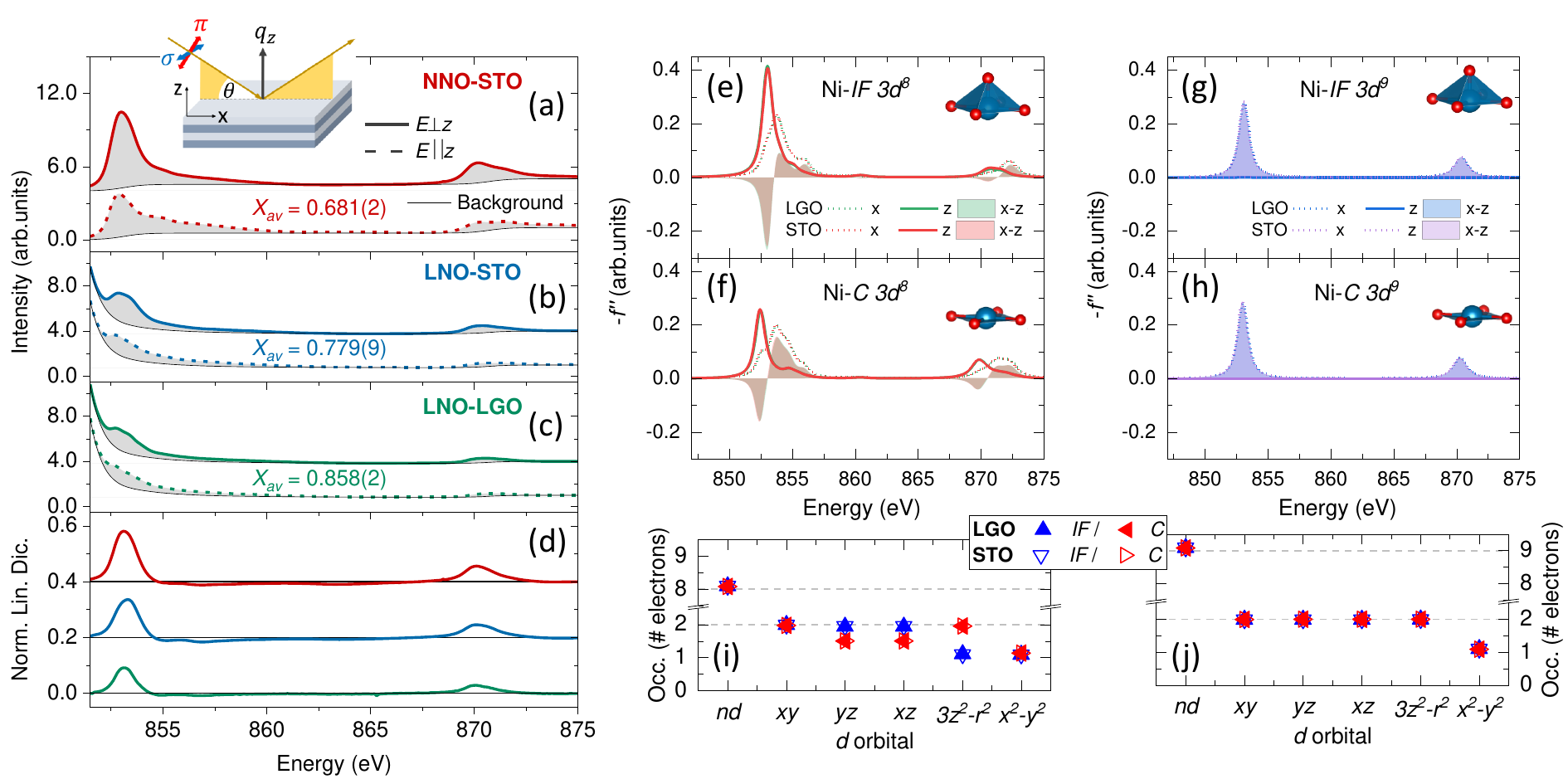}
\caption{(a-d) Experimental XAS measured with linearly polarized x-rays at room temperature for reduced (a) NNO-STO, (b) LNO-STO, and (c) LNO-LGO samples across the Ni-$L_{3,2}$ absorption edges. The solid lines in the top panels show spectra with $E\perp c$ and dashed lines with $E\parallel c$ polarization. The data for $E\perp c$ polarization are vertically shifted for clarity. (d) Normalized linear dichroism defined as $(I_{E\perp c} - I_{E\parallel c})/(2I'_{E\perp c} + I'_{E\parallel c})/3$ for all samples. The region of integration for calculating the $e_g$ hole ratio $X_{\text{av}}$ is marked in gray. The data have been vertically offset with respect to each other for clarity. The solid black lines in (a)-(c) represent the background subtracted (La-line and edge jumps). (e-h) Calculated spectra for the imaginary part of the scattering factor $f''\propto {\rm Energy} \cdot I_{\tiny XAS}$ for (e,f) Ni$^{2+}$ ($d^8$) and (g,h) Ni$^{1+}$ ($d^9$) in square-planar (C) and pyramidal (IF) ligand fields obtained from tight-binding fits to the layer-resolved DFT band structure. Additional parameters such as the Hubbard $U$, the charge transfer energy $\Delta$, and the Hartree Fock parameter are listed in Tab.~\ref{Tab_Cluster}. The energies were shifted by 845.6~eV (846.0~eV) for $d^8$ and by 845.0~eV (845.4~eV) for $d^9$ and LGO (STO) ligand fields, respectively. (i,j) Nickel $3d$-orbital occupations obtained from the different ligand-field cluster calculations.}\label{Fig_XAS_Cluster}
\end{figure*}

The samples were structurally characterized by x-ray diffraction (Appendix Fig.~\ref{Fig_XRD}). DC resistance measurements were carried out in van-der-Pauw geometry in a temperature range of $2$ - $300$\,K (Fig.~\ref{Fig_Resistivity}). X-ray absorption spectroscopy (XAS) and x-ray resonant reflectivity (XRR) measurements using linearly polarized soft x-rays ($\sigma$ and $\pi$ polarization) were performed on all samples at the UE46 PGM-1 beamline of the BESSY-II synchrotron at the Helmholtz-Zentrum Berlin. The scattering geometry is shown in the inset of Fig.~\ref{Fig_XAS_Cluster}(a). XAS was measured in total electron yield at incident angles of $\theta$ = 30\,$^{\circ}$, 45\,$^{\circ}$ (Fig.~\ref{Fig_XAS_Cluster}(a-d)). To obtain $E\parallel z$ and $E\perp z$ spectra we used the formulas $I_{E\perp z} = I_{\sigma}$ and $I_{\pi}(\theta) = I_{E\perp z}\textrm{cos}^2\theta + I_{E\parallel z}\textrm{sin}^2\theta$ \cite{Stoehr2006}.

To fit the reflectivity data, we used the software tool REMAGX \cite{Macke2014,Macke20141}. The structural model of each superlattice was determined by fitting simultaneously at least one off-resonance (10~keV) and two on-resonance (Ni-$L_{3,2}$) curves. Figure~\ref{Fig_XRR_q} in the Appendix shows the three sets of data of the NNO-STO, LNO-STO, and LNO-LGO superlattices. The black lines represent the fits for $\sigma$ and $\pi$ polarization of the incident x-rays. To fit the structural parameters (Appendix Tab.~\ref{Tab_XRRpara}) we used polarization-averaged optical constants, i.e.\ scalar scattering factors, and fixed them for the next analysis step when simulating the dichroic energy-dependent reflectivity. The $q_z$-values selected to perform the energy-dependent measurements shown in Fig.~\ref{Fig_XXR_NiL} correspond to the (002) and (003) superlattice reflections in NNO-STO and LNO-LGO. Since for LNO-STO the (003) is not well resolved, we chose to fit the (001) and (002) data instead (see also Ref.~\onlinecite{Ortiz2022}).

The ligand-field cluster calculations of the spectra used to analyze the layer-resolved results from the XRR modeling were performed with QUANTY \cite{Haverkort2012,Lu2014,Haverkort2014} with ligand-field parameters determined from DFT (see the next paragraph). The spectra for different fillings of Ni-$d$ states were calculated with the same input files by changing only the nominal number of Ni-$3d$ electrons from 8 to 9 and by replacing the corresponding Slater integrals and spin-orbit coupling parameters which were obtained from atomic Hartree-Fock calculations (see Appendix Tab.~\ref{Tab_Cluster}). We calculated the spectra for the different ligand fields in the Ni layer at the interface (IF) and in the most central layers of the stack (C) and for different interlayer materials STO and LGO (Fig.~\ref{Fig_XAS_Cluster}(e-h)). The room temperature spectra were calculated as sums over the different states, weighted by the Boltzmann occupation factor. We also varied the parameter $U$ from 4 to 7~eV and $\Delta$ from 3.5 to 9~eV and found no relevant influence. The results shown in Fig.~\ref{Fig_XAS_Cluster}(e-h) were obtained for $U_{dd}=6$~eV and $\Delta=4.5$~eV. These values are close to those reported in Ref.~\cite{LaBollita2022}. Only the relative energy axis is given for the calculated spectra, but the absolute energy values depend on the energy of the $2p$ core levels and must be shifted. We have shifted the spectra by 845.6~eV + $\Delta_{ABO_3}$ - $\Delta_{d^9}$, where $\Delta_{ABO_3}$ = 0.42~eV is a shift accounting for different STO and LGO ligand-field energies and $\Delta_{d^9}$ = 0.6~eV is the core level shift for $d^8$ to $d^9$, which we estimated by comparing the averaged IF-$d^8$ and C-$d^9$ spectra with the absorption spectra in Fig.~\ref{Fig_XAS_Cluster}(a-d).

The numerical, single particle part of the ligand-field Hamiltonian used in the cluster calculations was obtained from the DFT band structures considering the superlattice structure and downfolding to a set of localized Ni-$3d$ and O-$2p$ Wannier orbitals. The DFT calculations to obtain the electronic ground state were carried out using the Quantum ESPRESSO suite \cite{giannozzi2009quantum,giannozzi2017advanced} and subsequently the plane-wave basis was projected to a basis of maximally-localized Wannier functions with the Wannier90 program \cite{pizzi2020wannier90}. Optimized norm-conserving pseudopotentials \cite{hamann2013optimized,van2018pseudodojo} were used with an energy cutoff of 110~Ry and the exchange-correlation interaction was approximated by the Perdew-Burke-Ernzerhof functional (PBE). Brillouin zone integration for the calculation of the electronic ground state was carried out on a $6\times 6\times 1$ electron momentum grid, where a Gaussian smearing of 0.02~Ry was included.

The $z$-components of the atomic positions were relaxed until the residual forces were lower than 0.036~eV/\AA, while the $x$ and $y$ positions were kept fixed by the tetragonal symmetry. The lattice constants of the superlattice cell were set to the experimental values determined by x-ray diffraction (Appendix Fig.~\ref{Fig_XRD}): The in-plane lattice constant $a=3.903~$\AA\ is equal to the one of the STO-(001) substrate, while the out-of-plane lattice constant was $c=42.8$~\AA\ for RNO-STO and $c=43.38$~\AA\ for LNO-LGO, respectively.

For the projection onto the Wannier basis atom-centered Ni-$d$ and O-$p$ states were used as initial projections and the disentanglement routine of Wannier90 was utilized. Here the inner (frozen) window was chosen from about -10~eV to 0~eV (w.r.t.\ the Fermi level), corresponding to the manifold of bands, which is exclusively of Ni-$d$ and O-$p$ character. The outer window was extended from -10~eV to +10~eV to incorporate the Ni-$d$ and O-$p$ character of the conduction bands.

\section{Results}
\subsection{Resistivity}

We start our discussion by analyzing the transport data shown in Fig.~\ref{Fig_Resistivity}. Superlattices with pristine $R$NiO$_3$ stacks show bulk-like resistivity behavior, i.e.~ a metallic temperature dependence of resistivity for LaNiO$_3$ and a first order metal-insulator transition (MIT) around T$_{\textrm{\scriptsize MIT}} \approx 150$\,K for NdNiO$_3$ \cite{Breckenfeld2014,Allen2015,Yao2016} (Fig.~\ref{Fig_Resistivity}(a)).

Annealing with CaH$_2$ modifies the resistivity behavior of all superlattices. While the LNO-LGO superlattice shows a semiconducting temperature dependence, as already reported in Ref.~\cite{Ortiz2021}, the LNO-STO and NNO-STO superlattices show metallic behavior with a weak upturn at low temperatures. This difference could potentially stem from electronic reconstructions resulting from the polar discontinuity of the $R$NO-STO interface, leading to a nominal charge sequence of SrO$^{0}$-TiO$_2^{0}$-$R$$^{3+}$-NiO$_2$$^{-3}$ for the reduced superlattices. These polar discontinuity models assume an exact oxygen stoichiometry in the layers, but it is precisely this stoichiometry that is adjusted in the topotactic reduction process. We have identified an interface structure of the LNO-LGO superlattices in which the apical oxygen is not removed in the layers next to LGO \cite{Ortiz2021}, thus naturally resolving the polar discontinuity structurally \cite{Bernardini2022}. This interface structure was also recently confirmed by STEM imaging \cite{Yang2023}. The remaining oxygen at the interface donates holes to the nickelate layers. Assuming homogeneous hole doping across the nickelate layer, the doping level $c$ depends on the layer thickness, i.e.\ the number $m$ of NNO u.c.\ stack along the growing direction, as pointed out in the introduction. Therefore, we have examined the resistance when varying the NNO stack thickness from $m=6$ to 12~u.c. (Fig.~\ref{Fig_Resistivity}(c)). We observe a metallic temperature dependence of resistivity, independent of $m$, however a variation in the conductivity values above 50~K. The inset displays the residual resistivity ($\rho_{\rm res}$), which was obtained from extrapolating the high-temperature resistivity. The star symbols are $T=20$~K resistivity values that were measured on Nd$_{1-x}$Sr$_{x}$NiO$_2$ thin films in Ref.~\onlinecite{Li2020}. The horizontal gray line indicated the quantum sheet resistance per NiO$_2$ two-dimensional plane. Superconducting samples in Ref.~\onlinecite{Li2020} showed normal state resistivity at 20~K below this value of 850–880~$\mu\Omega$cm across the doping series. In a recent publication documenting superconductivity in undoped NNO films \cite{Parzyck2024}, the authors find superconductivity in only about half of the samples that have $\rho_{\rm res}$ below this value. In other words, the high quality (phase purity) characterized by a low value of $\rho_{\rm res}$ seems to be a necessary but not sufficient condition for the occurrence of superconductivity. The values of $\rho_{\rm res}$ of all superlattice samples except $m=6$ fall below this line. We thus conclude that the quality and size of the disorder in the nickelate stacks of our superlattices is comparable to those of superconducting thin films in Refs.~\cite{Li2020, Parzyck2024}.

We will first determine the layer-dependent variation of the electronic structure within the nickelate stacks and for various interlayer materials before returning to the altered resistivity behavior. The layer-selectivity of the reduction process has been tested akin to the LGO SL studied in Ref.~\cite{Ortiz2021} by analyzing the Ti-$L$ XAS (Appendix Fig.~\ref{Fig_XAS_TiL}), where we found no indications for reduction-related changes in STO interlayers. All spectra are characteristic of a dominant Ti$^{4+}$ valence state and we thus conclude that the reduction process is layer selective, as observed in related superlattice systems \cite{Matsumoto2011}.

\subsection{X-ray absorption}

The layer-averaged, local electronic structure of the nickelate stacks is reflected in the Ni $L$-edge XAS (Fig.~\ref{Fig_XAS_Cluster}(a-d)). The data for all reduced superlattices exhibit qualitatively similar spectral shapes, where the main absorption peaks at $L_3$ and $L_2$ are shifted by about 1~eV towards lower energy as compared to their pristine counterparts \cite{Ortiz2021}, indicating a lowering of the layer-averaged nickel valence state. The reduced samples show positive dichroism indicating more holes in the $d_{x^2-y^2}$ orbital, compared to the $d_{3z^2-\textbf{r}^2}$ orbital. To quantitatively estimate the ratio of holes in the Ni-$e_{\textrm{\scriptsize g}}$ orbitals, we used the following equation, which is derived from the sum rules \cite{Benckiser2011}:
\begin{equation}
    X_{\textrm{\scriptsize av}} = \frac{h_{3z^2-\textbf{r}^2}}{h_{x^2-y^2}}=\frac{3I'_{E\parallel\,z}}{4I'_{E\perp\,z}-I'_{E\parallel\,z}} \label{hole},
\end{equation}
where $h_{x^2-y^2}$ and $h_{3z^2-\textbf{r}^2}$ denote the hole occupation in the $d_{x^2-y^2}$ and $d_{3z^2-\textbf{r}^2}$ orbitals, respectively, and $I'_{E\perp\,z, E\parallel\,z} = \int_{L_{3,2}} I_{(E\perp\,z, E\parallel\,z)}(E)dE$ is the integrated intensity over the entire Ni-$L_{3,2}$ absorption edge. We note that for calculating the $X_\mathrm{{av}}$ values, the La-$M_4$ lines and edge jumps (solid black lines in Fig.~\ref{Fig_XAS_Cluster} (a)-(c)) had to be subtracted before integration and the errors reflect the estimated uncertainty of this procedure. The $X_\mathrm{{av}}$ values of the reduced samples indicate an overall preferential hole occupation of the $d_{x^2-y^2}$ orbitals with values of $X_\mathrm{{av}}$ clearly smaller than unity. Moreover, comparing the different superlattices, the preferred hole occupation of the $d_{x^2-y^2}$ orbital increases from LNO-LGO to LNO-STO to NNO-STO.

In our previous study \cite{Ortiz2021} the DFT+DMFT results indicated that both the ligand field and the $d$ filling are different in the interface (IF) and central (C) layers of the $R$NO stacks in the LNO-LGO superlattice. Since the spectra and in particular the linear dichroism at the Ni-$L$ edge spectra are very different, as our cluster calculations in Fig.~\ref{Fig_XAS_Cluster}(e-h) show, it is imperative to determine the layer-resolved results within the $R$NO layer stacks. For this purpose, we have carried out XRR measurements.

\subsection{X-ray resonant reflectivity}\label{XRRr}

\begin{figure*}[tb]
\center\includegraphics[width=0.98\linewidth]{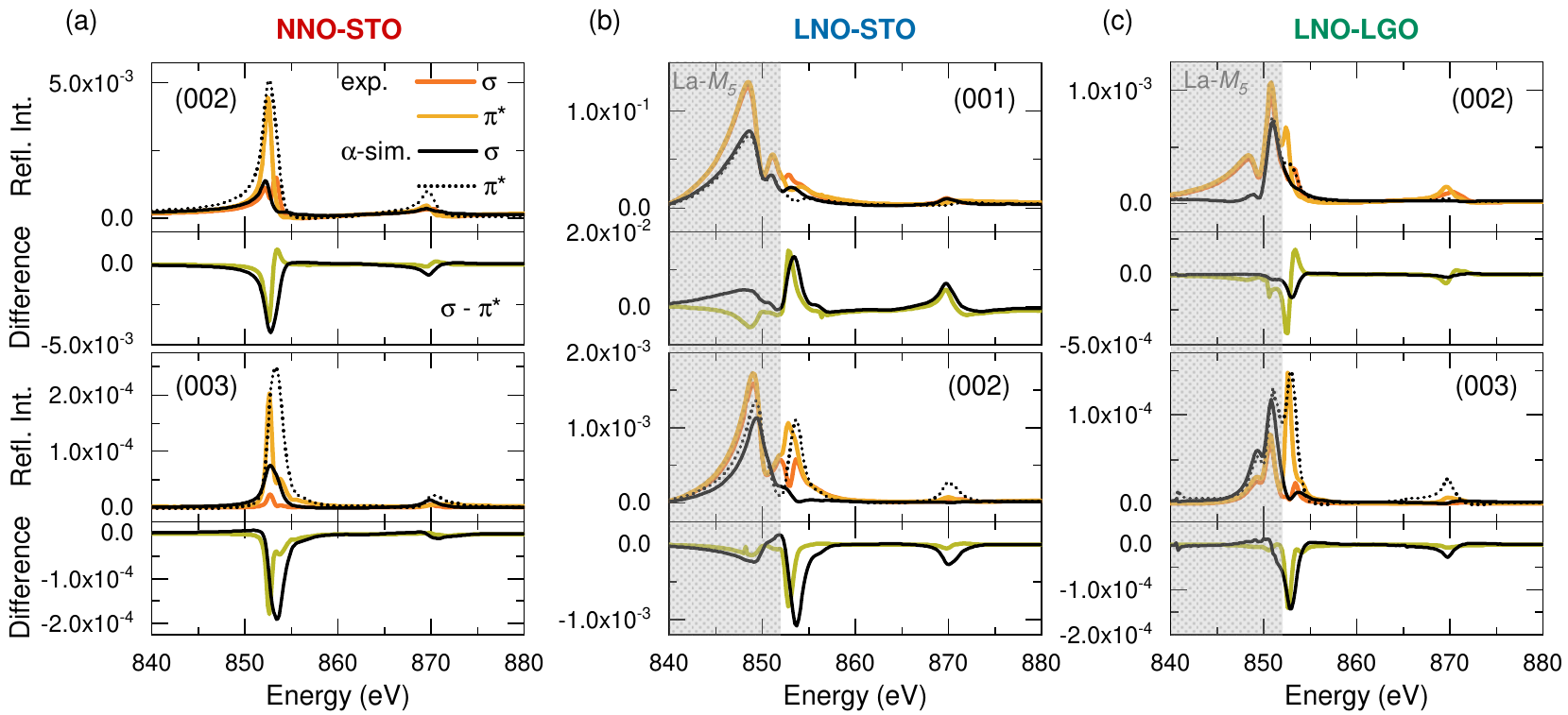}
\caption{Reflectivity constant-q$_z$ energy scans for the reduced (a) NNO-STO, (b) LNO-STO, and (c) LNO-LGO superlattices. Each panel shows the measurements at different $(00L)$ reflections (colored lines) and the simulated spectra obtained by redistributing the dichroism with a parameter $\alpha$ (black lines). The $\pi^*$ spectra correspond to $\pi_{pol}\times [\cos(2\theta)^{-2}]$, which is a correction to the intensity minimum in the vicinity of the Brewster angle. The shaded areas correspond to the region dominated by the La-$M_4$ contribution.}\label{Fig_XXR_NiL}
\end{figure*}

In the first analysis step to determine layer-resolved spectra, we measured the reflectivity as a function of $q_z$ at fixed energies at the Ni-$L_{3,2}$ edges and at high (off-resonant) energies around 10\,keV and simultaneously fitted these curves to obtain a structural model (Appendix Fig.~\ref{Fig_XRR_q} and Tab.~\ref{Tab_XRRpara}).

\begin{figure*}[tb]
\center\includegraphics[width=0.99\linewidth]{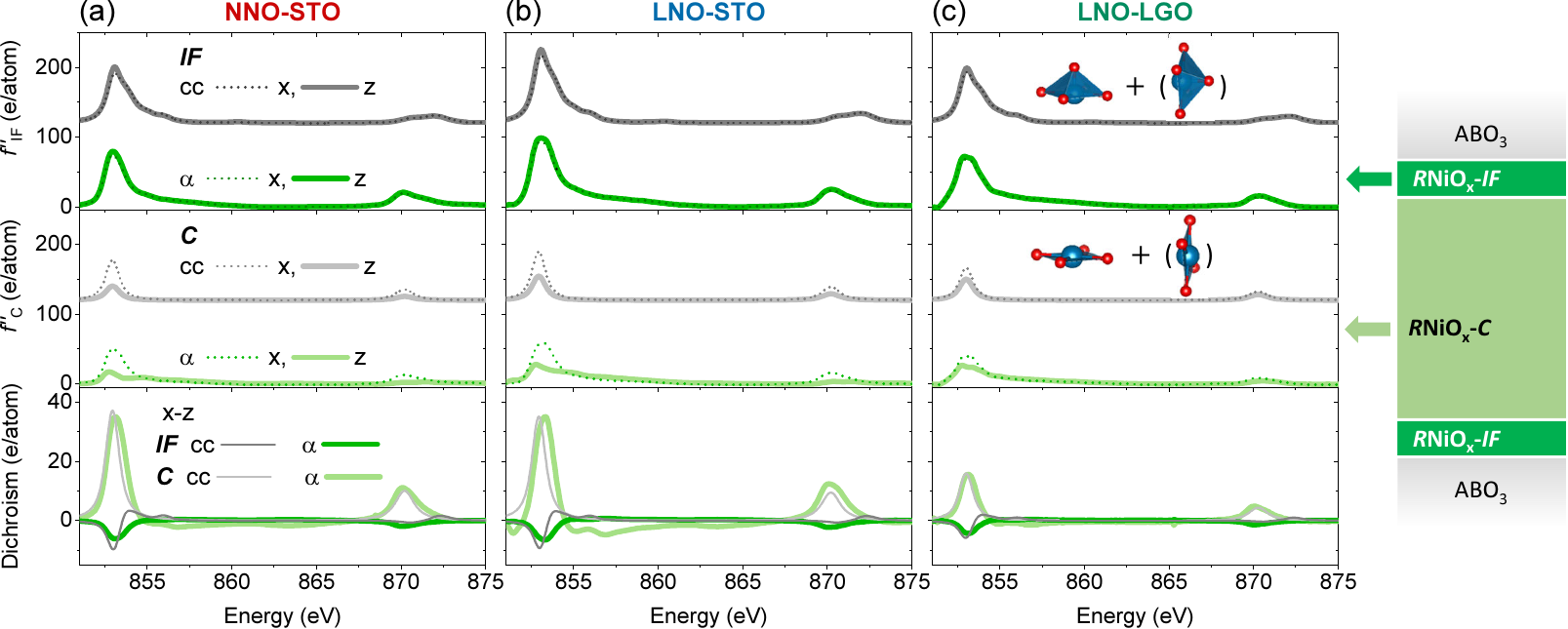}
\caption{Layer-resolved scattering factor $f$'' at the Ni-$L_{3,2}$ edges for three types of reduced superlattices: (a) NNO-STO, (b) LNO-STO, and (c) LNO-LGO. The results of the two simulation methods - dichroism distributed by the parameter $\alpha$ (green lines) and cluster calculations (cc; gray lines) - which were renormalized to $f''_{IF} = 2\cdot f''_{C}$ to satisfy the sum rules. The curves in the top panels show the spectra for the interfacial layers (IF), the curves in the middle panels correspond to the central layers (C), and the bottom panels show the linear dichroism ($f''_{x} - f''_{z}$) of the C and IF layers from both methods.} \label{Fig_Layer_res_spectra}
\end{figure*}

In the second analysis step, we simulated the energy-dependent reflectivity at constant $q_z$ and compared the results to the experimental curves. The experimental data in Fig.~\ref{Fig_XXR_NiL} (colored curves) show that the spectral shape strongly depends on the value of $q_z$ with a pronounced polarization dependence at the Ni-$L$ edge (see the dichroic difference spectra in the respective panels below). Note that the spectra of LNO-STO and LNO-LGO also overlap with the La-$M_4$ absorption line (shaded region), which shows negligible dichroism. To model the spectra, we subdivided the nickelate layer stacks into interfacial (IF) and central (C) layers and allowed different complex scattering tensors $f_{\textrm{\tiny IF}}$ and $f_{\textrm{\tiny C}}$ in these layers (see sketch in Fig.~\ref{Fig_Layer_res_spectra} on the right). In a simplified model for the particular 8/4 superlattice, assuming equal out-of-plane lattice spacing and no roughness, the intensity at the $(002)$ and $(003)$ superlattice reflections is sensitive to $(f_{\textrm{\tiny ABO}}-f_{\textrm{\tiny IF}})$ and $(f_{\textrm{\tiny IF}}-f_{\textrm{\tiny C}})$, respectively, while the (001) superlattice reflection arises from contributions of all layers $( f_{\textrm{\tiny ABO}}, f_{\textrm{\tiny C}}$, and $f_{\textrm{\tiny IF}})$. Therefore, energy- and polarization-dependent reflectivity measurements with a fixed-momentum transfer at (003) are particularly sensitive to a modulation of the electronic structure within the nickelate stack in the superlattices.

We employed an optical model, where the scattered intensities in each layer are described through a dielectric tensor $\hat{\varepsilon}$ of tetragonal symmetry. To obtain the energy- and layer-dependent entries for $\hat{\varepsilon}(E)$, optical constants along in-plane and out-of-plane directions are required \cite{Benckiser2011}. To simulate the energy-dependent reflectivity at constant $q_z$, we started with the same approach as in Ref.~\onlinecite{Benckiser2011}, where the entries of $\hat{\varepsilon}$ of IF and C layers were obtained from the measured averaged linear dichroism in XAS, parameterized by a factor $\alpha$ that redistributes the dichroism between the C and IF nickelate layers. The corresponding dielectric tensor entries for this case are given by:
\begin{equation}
\left(\varepsilon_{ \textrm{\tiny C(IF)}}\right)^{jj} = \left(1 \mp p_{\textrm{\tiny IF(C)}}\alpha\right)\varepsilon^{jj} \pm p_{\textrm{\tiny IF(C)}}\alpha\, \varepsilon_{\textrm{\tiny cubic}}, \label{tensor}
\end{equation}
with upper signs for C and lower ones for IF layers. The complex entries of the dielectric tensor $\varepsilon^{jj}=\varepsilon^{jj}_1 + i\varepsilon^{jj}_2$ $(j=x,z)$ are related to the scattering factors, without an index for the one measured by XAS, and with index IF and C related to $f_{\textrm{\tiny IF}}$ and $f_{\textrm{\tiny C}}$, $\varepsilon_{\textrm{\tiny cubic}} = \frac{1}{3}(2\varepsilon^{xx}+\varepsilon^{zz})$ and $p_{\textrm{\tiny IF(C)}}$ are the numbers of unit cells of IF (C) layers. The parameter range of $\alpha$ in Ref.~\onlinecite{Benckiser2011} was extended to $\alpha\in[-(p_{\textrm{\tiny C}})^{-1}, (p_{\textrm{\tiny IF}})^{-1}]$ in order to allow for the redistribution of dichroism with opposite sign in C and IF layers, and to account for the asymmetry in the number of layers of each material in the superlattice, i.e.\ 8 vs.\ 4~u.c. Three cases can be distinguished: (i) for $\alpha = 0$, central and interface layers exhibit the same dichroism as is measured in XAS, i.e.\ $\varepsilon_{ \textrm{\tiny C}}=\varepsilon_{ \textrm{\tiny IF}}=\varepsilon$. (ii) In the case of $\alpha > 0$, the central layers contribute less to the total dichroism than the interface layers, and (iii) for $\alpha < 0$ it is the opposite case. For all reduced samples, we obtained negative values of $\alpha$, indicating a preferential hole-occupation of the $d_{x^2 - y^2}$ orbital of Ni-$3d^9$. The best model effectively reproduced the linear dichroic reflectivity (solid and dotted black lines in Fig.~\ref{Fig_XXR_NiL}).

From the model obtained by varying the $\alpha$ parameter, we can determine the layer-resolved imaginary part of the scattering factor, which is proportional to the x-ray absorption intensity via the optical theorem: $I_{\tiny XAS}(E)\propto \frac{1}{E}f''(E)$. The results are shown in Fig.~\ref{Fig_Layer_res_spectra} (solid/dashed green curves). Most strikingly, we obtain very distinct dichroic spectra for C and IF layers: The dichroism of the C layer is large and positive, clearly providing the main contribution to the layer-averaged XAS and qualitatively similar to what is expected for the Ni-$3d^9$ configuration (see Fig.~\ref{Fig_XAS_Cluster}(e)). On the contrary, the dichroism of the IF layer is more subtle and derivative-like across the Ni-$L_{3,2}$ energies, reminiscent of high-spin Ni-$3d^8$ \cite{Haverkort2004}.

Ligand-field cluster calculations were performed to explicitly account for the possible different orbital occupations in the C and IF layers and the resulting distinct dichroism. We have performed four calculations, for the square-pyramidal (IF) and for the square-planar (C) ligand field with parameters determined from DFT calculations, each with a nominal $3d$-filling of $nd = 8$ (Ni$^{2+}$) and 9 (Ni$^{1+}$) electrons, respectively (Fig.~\ref{Fig_XAS_Cluster}(e-j) and Appendix). The results show that the negative dichroism in the IF square-pyramidal layers implies a Ni-$d^8$ configuration, while the positive dichroism in the C layers with the one-peak spectrum and non-zero integral indicates Ni-$d^9$. All spectra are only insignificantly dependent on ligand field changes due to STO or LGO interlayers. The different Ni valence in the IF and C layers was also found by analyzing the $L_3/L_2$ intensity ratios in STEM-EELS data measured on the NNO-STO superlattice \cite{Yang2023}, however a quantitative evaluation of the layer-specific orbital occupations is not possible in STEM.

The Ni-$d$ orbital occupations obtained from the ligand-field calculations are plotted in Fig.~\ref{Fig_XAS_Cluster}(i,j). We compare the total Ni-$d$ electron occupation $nd$ and the occupations of the individual five $3d$ orbitals for the Ni-IF vs.\ Ni-C ligand fields and different nominal filling of $d^8$ and $d^9$. The hybridization with the oxygen ligand orbitals leads to a similarly increased electron filling $nd$ for $d^8$ and $d^9$. For all cases, we find no relevant difference between LGO and STO interlayers. For the $d^9$ configuration, while there is also no significant effect of the ligand field, i.e.\ no difference for Ni-IF and Ni-C in the spectra or orbital occupations (there is always one hole in the $x^2-y^2$ orbital), there is a clear difference for square pyramidal Ni-IF vs. square-planar Ni-C ligand field. While for Ni-IF all $t_{2g}$ orbitals are full, and two Hund's-coupled spins are in the $e_g$ orbitals, for Ni-C there is one hole in the $x^2-y^2$ and the other one is in the degenerate $xz/yz$ orbitals. In other words, hole doping the $d^9$ configuration leads to partial removal of electrons in the $xz/yz$ orbitals. For the $d^8$ configuration, we find that the orbital polarization, given by the integrated area of the dichroic spectrum $x-z$, is significantly larger for square pyramidal Ni-IF than for square planar coordinated Ni-C.

The XRR data in combination with cluster calculations allow for such a quantitative comparison. First we ensure that the $f_{\alpha}^{''}(E)$-spectra obtained by varying the $\alpha$-parameter fulfill the sum rule and reflect the difference in valence states, i.e.\ the integrated areas of the polarization-averaged spectra of IF-$d^8$ (2 holes) and C-$d^9$ (1 hole) must have a 2:1 ratio. The direct comparison of the cluster calculation results from $f^{''}_{cc}(E)$ and the properly normalized $f_{\alpha}^{''}(E)$ shows that the dichroism in the experimentally obtained XRR spectra is clearly smaller than the one predicted by the cluster calculations (Fig.~\ref{Fig_XAS_Cluster}(e-j)). A possible explanation for this phenomenon is that oxygen ligands are removed unevenly during the reduction process, resulting in cases where basal rather than apical oxygen is removed from certain Ni sites. Such an alternative, vertically aligned ligand field leads to spectra with reversed dichroism, i.e.\ $x=z$ and $z=x$, without a change in the $d$ filling. By mixing the spectra of both polarizations so that they match the ones of $f_{C}^{''(\alpha)}(E)$ and $f_{IF}^{''(\alpha)}(E)$, respectively, we can estimate the percentage of reverse-oriented sites in IF and C layers:
\begin{eqnarray}
f_{IF\,x(z)}^{''(\alpha)}(E)=N f^{''(cc)}_{IF\,x(z)} + (1-N) f^{''(cc)}_{IF\,z(x)}\\
f_{C\,x(z)}^{''(\alpha)}(E)=M f^{''(cc)}_{C\,x(z)} + (1-M) f^{''(cc)}_{C\,z(x)}
\end{eqnarray}
The results represented by the solid/dashed gray curves in Fig.~\ref{Fig_XXR_NiL} reveal a substantial admixture of the respective other orientation for both, IF and C layers, which depends on the interlayer material and explains the trends observed in the quantitative analysis of the layer-averaged linear dichroism measured in XAS (Fig.~\ref{Fig_XAS_Cluster}(a-c)). We found that for all superlattices the interface layers have almost equal contributions of in-plane and out-of-plane oriented square-pyramidal configurations ($\approx 45-55\%$). In contrast, the main contribution for the central layers is the out-of-plane oriented square-planar configuration. The ratio is around 70/30\,\% in the superlattices with STO and 60/40\,\% in the superlattice with LGO. Thus, superlattices with STO show less disorder on the anion lattice in the central layers and a preference for apical oxygen removal compared to the sample with LGO interlayers. This behavior correlates with the electrical transport measurements (Fig.~\ref{Fig_Resistivity}(b)), where NNO-STO and LNO-STO superlattices with less disorder show higher conductivity.

In the last step, we cross check our analysis by using the spectra obtained from the cluster calculation with subsequent polarization mixing as input for the simulation of the constant-$q$ reflectivity data (Fig.~\ref{Fig_XRR_NiL_cluster}). The spectra obtained with this input agree well with the experimental data. While the line shape of the individual spectra for $\sigma$ and $\pi$ polarization is slightly less well reproduced than with the measured spectra as input (Fig.~\ref{Fig_XXR_NiL}), the dichroic difference spectrum is better described, because the energy shift between Ni$^{1+}$ and Ni$^{2+}$ could be taken into account for the cluster calculation input. Alternative models with mixed Ni$^{1+}$/Ni$^{2+}$ valence in interfacial and central layers all resulted in poor agreement with the experimental data (see Appendices G and H of Ref.~\onlinecite{Ortiz2022}).

Finally, we varied the relative layer thicknesses of interface and central layers ($p_{\textrm{\tiny IF}}+p_{\textrm{\tiny C}}=8$ u.c.) and found that the best match is obtained when $(p_{\textrm{\tiny IF}}/p_{\textrm{\tiny C}})_{\textrm{\tiny LNO-LGO}} = 2/6$ and $(p_{\textrm{\tiny IF}}/p_{\textrm{\tiny C}})_{\textrm{\tiny LNO-STO}} = (p_{\textrm{\tiny IF}}/p_{\textrm{\tiny C}})_{\textrm{\tiny NNO-STO}} = 3/5$. These results show that in the LNO-LGO samples, the top and bottom interfacial layers are about one u.c.\ thick, while in NNO-STO and LNO-STO they are slightly thicker, i.e.\ 1.5 ~u.c.\ thick. This implies that oxygen removal is less favorable in the immediate vicinity of STO and that the presence of Ni$^{2+}$ extends over a larger range from the interface into the nickelate layer stack. One possible explanation for this is the formation of a mixed $R$(Ti$_{0.5}$Ni$_{0.5}$)O$_3$ layer, followed by LaO-NiO$_2$-La atomic layers (all with Ni$^{2+}$) as recently suggested by a combined STEM and DFT$+U$ study \cite{Goodge2023}.
\section{Conclusions}

The layer-resolved spectra obtained from XRR, in combination with ligand-field cluster calculations on superlattices with $m$=8, provide information about the local Ni electronic configuration and orbital occupations at the interface and in the central part of the nickelate layer stacks. Our results show that in contrast to superlattices with LGO interlayers, the superlattices with STO exhibit metallic conductivity after reduction akin to the undoped infinite-layer compounds \cite{Li2019}. The depth-resolved spectra confirm the previously found residual oxygen also in the interface layers and that the holes provided by this residual oxygen are distributed unevenly across the nickelate slabs. While at the interfaces square-pyramidally coordinated Ni$^{2+}$ is present, the central part of the nickelate stack contains square-planar coordinated Ni$^{1+}$. Consistent with the observed difference in resistivity, this spatial distribution depends on the interlayer material, and we observe a more extended interlayer in superlattices with STO. The quantitative analysis of the XRR dichroism shows that there is a significant fraction of disoriented coordination polyhedra in which the basal oxygen is removed instead of the apical one. Both, the Ni$^{2+}$O$_5$ pyramids in the interface layers as well as the Ni$^{1+}$O$_4$ plaquettes in the central layers of the nickelate stacks are oriented partly in-plane and partly out-of-plane and the ratio of the latter also depends on the interlayer material. The superlattices with STO interlayers clearly show a favored in-plane orientation of the plaquettes. Since the numbers in both STO superlattices are the same within our error bars, we conclude that the choice of rare-earth ions $R$ has no decisive influence. However, the possible influence of the STO layer (as a superlattice spacer or as a capping layer on a film \cite{Raji2023}) on the residual oxygen and the orientation of the oxygen sublattice provides an important clue regarding the origin of some contradictory experimental results. In particular, it will be important to determine what controls the oxygen disorder and domain size of the out-of-plane aligned plaquettes. The results of the present work provide information on the disorder in the nickelate layer stacks in superlattices over large spatial volumes of the sample, corresponding to that of our x-ray illuminated region with a diameter of $\sim 500~\mu$m times the total superlattice thickness $\sim 35$~nm. Such information is complementary to STEM(-EELS) measurements \cite{Yang2023}, which can only determine valence profiles averaged over small spatial areas and over the specimen thickness, and give only qualitative information about disorder in the oxygen sublattice. The four-dimensional STEM study on one of the NNO-STO superlattices \cite{Yang2023} provides a lower limit, as it detected no significant disorder in the oxygen lattice on a $\sim$~20~nm lateral scale. Since this length scale can also depend on the composition, systematic and quantitative investigations of oxygen disorder are an important task for future studies.

\begin{acknowledgments}
The authors gratefully acknowledge financial support by the Center for Integrated Quantum Science and Technology (IQ$^{\rm ST}$). We further acknowledge the scientific support and HPC resources provided by the Erlangen National High Performance Computing Center (NHR@FAU) of the Friedrich-Alexander-Universit\"{a}t Erlangen-N\"{u}rnberg (FAU). The hardware is funded by the German Research Foundation (DFG). We thank Helmholtz-Zentrum Berlin for the allocation of synchrotron radiation beamtime at BESSY II.
The Institute for Beam Physics and Technology (IBPT) at the Karlsruhe Institute of Technology (KIT) is accredited for the operation of the storage ring Karlsruhe Research Accelerator (KARA), and for provision of beamtime at the KIT light source. We thank P.~Puphal for helpful discussions.
\end{acknowledgments}

\appendix

\section{X-ray diffraction characterization}

\begin{figure}[h]
\center\includegraphics[width=0.99\columnwidth]{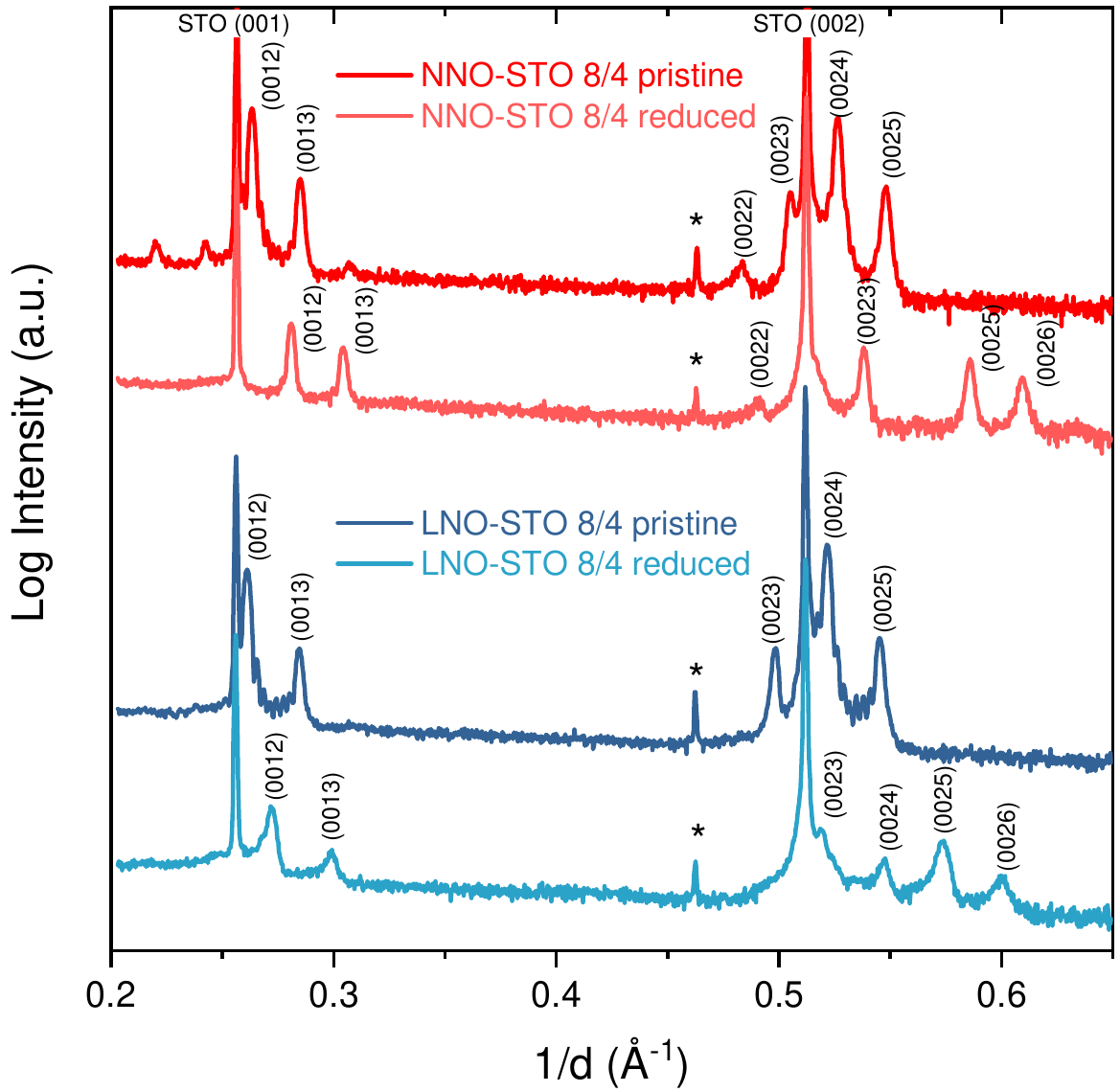}
\caption{XRD patterns measured on pieces of the same as-grown sample of pristine and reduced superlattices covering the (001) and (002) reflections of the cubic SrTiO$_3$ substrate. In each panel, the data for pristine and reduced samples have been translated vertically for clarity. The asterisks mark peaks
arising from the $K_{\beta1}$ reflection of the substrate peak.}\label{Fig_XRD}
\end{figure}

X-ray diffraction $\theta / 2\theta$ measurements were carried out with a custom-made four-circle single-crystal diffractometer, which is equipped with a Cu-$K_{\alpha1}$ source and a Dectris MYTHEN line detector. We observe the same intensity modulation in the superstructure peaks above STO(002) as reported for the LNO-LGO superlattice in \cite{Ortiz2021}, suggesting that square-pyramidal nickel coordination is also present at the interfaces with STO, consistent with the results of the XRR analysis.

\section{X-ray absorption at the Ti-\textit{L} edge}
\begin{figure}[h]
\center\includegraphics[width=0.95\columnwidth]{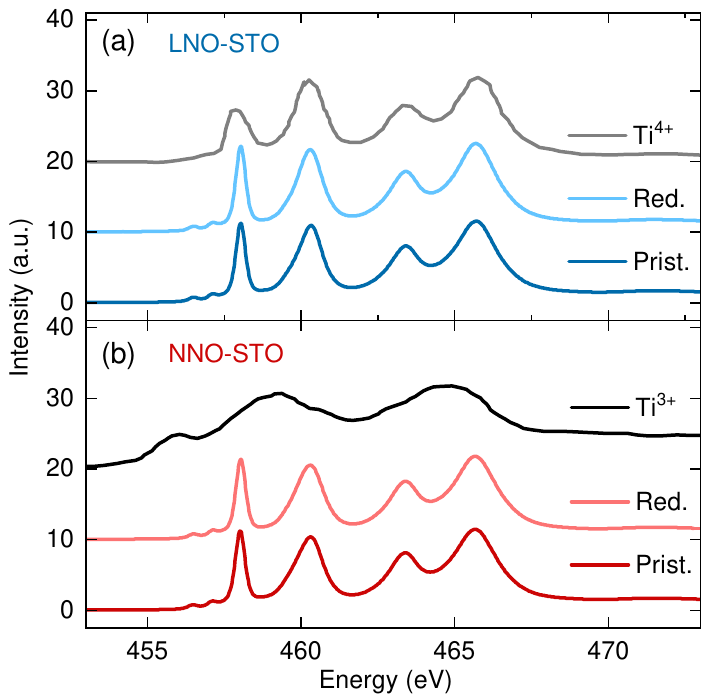}
\caption{Polarization averaged XAS at the Ti-$L_{3,2}$, measured in total electron yield for the pristine (dark colors) and reduced (light colors) LNO-STO (a) and NNO-STO samples (b). The reference data for Ti$^{4+}$ (grey line in (a)) and Ti$^{3+}$ (black line in (b)) were reproduced from Ref.~\onlinecite{Cao2016}. The data for the reduced samples and the reference spectra for Ti$^{4(3)+}$ were vertically translated for clarity.}\label{Fig_XAS_TiL}
\end{figure}

Figure~\ref{Fig_XAS_TiL} shows the unpolarized XAS at the Ti-$L$ edge.

\begin{figure*}[tb]
\center\includegraphics[width=0.98\linewidth]{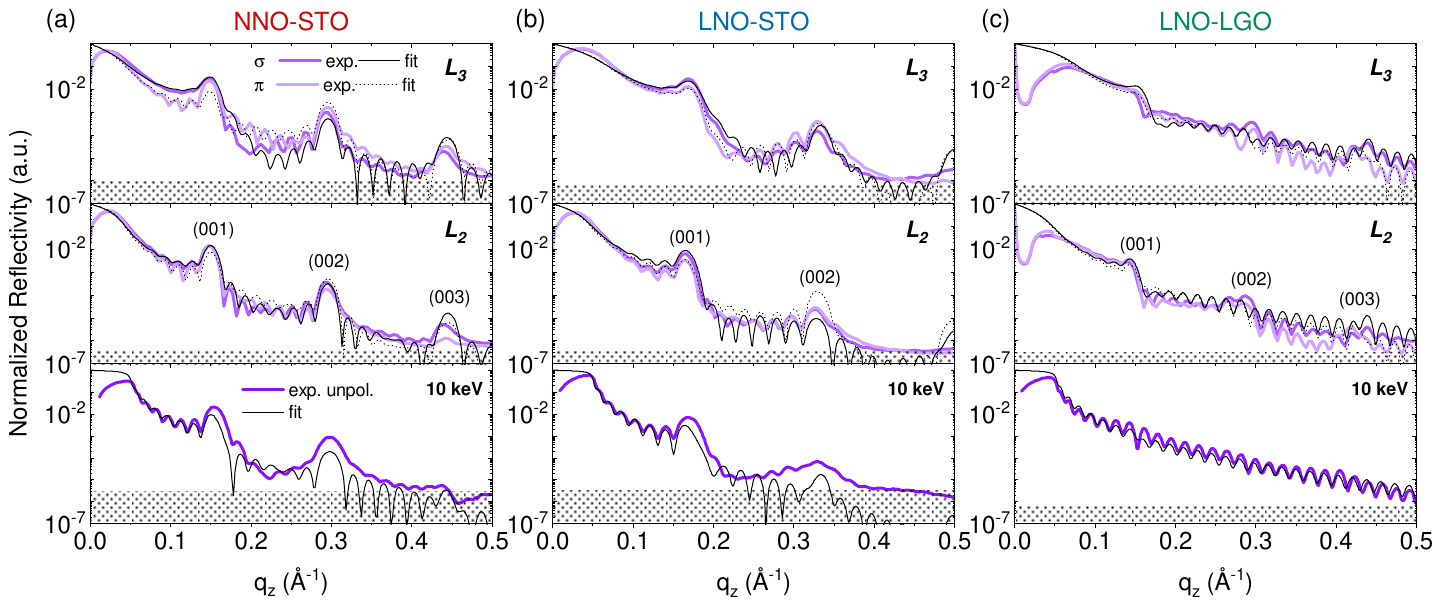}
\caption{Reflectivity as a function of $q_z$ for the reduced (a) NNO-STO, (b) LNO-STO, and (c) LNO-LGO superlattices. Each panel shows three curves corresponding to one off-resonant (bottom panel) and two resonant measurements at the Ni-$L$ edges (middle and top panel). Ni-L$_3$ edge energy values are 853.0\,eV (NNO-STO), 852.8\,eV (LNO-STO), and 853.4\,eV (LNO-LGO). Ni-L$_2$ edge energy values are 870.1\,eV (NNO-STO), 870.3\,eV (LNO-STO), and 870.6\,eV (LNO-LGO). In the case of the off-resonant measurements all data were collected with 10\,keV x-rays. The labels (001), (002), and (003) correspond to the superlattice reflections. The dashed areas in each panel represent the noise level of the measurement.}\label{Fig_XRR_q}
\end{figure*}

\section{Momentum dependent x-ray reflectivity}

The profiles resulting from these fits were obtained using the polarization-averaged optical constants (defined as $I_{\text{av}}$=(2$\cdot I_{E\perp z}+I_{E\parallel z}$)/3) measured in XAS. These optical constants were used to obtain scalar scattering factors introduced in a Parrat fitting routine \cite{Parrat1954}. The best agreement was obtained for a structural model where the first $R$NO slab at the substrate interface and the last ABO slab at the SL surface were allowed to have different thicknesses and roughnesses (see Supplementary Material Tab.~I). In the subsequent modeling of the dichroic reflectivity, the structural parameters were kept fixed.

\begin{table*}[bht]
\centering
\begin{adjustbox}{width=1\textwidth}
\setlength{\tabcolsep}{12pt}
\begin{tabular}{lccc|ccc}
\hline\hline
\multicolumn{1}{c}{\multirow{2}{*}{Fit Parameters}} & \multicolumn{3}{c|}{Thickness (\AA)}  & \multicolumn{3}{c}{Roughness (\AA)}   \\ \cline{2-7}
\multicolumn{1}{c}{}                                & NNO-STO     & LNO-STO     & LNO-LGO     & NNO-STO     & LNO-STO     & LNO-LGO     \\ \hline
7 layers (Nd, La)NiO$_{2+\delta}$                            & 27.4 & 24.0 & 27.6 & 3.4 & 4.8 & 7.6 \\
7 layers of STO (LGO)                            & 15.2 & 14.0 & 16.0 & 2.7 & 6.6 & 7.2 \\
STO  substrate                                 & $\infty$    & $\infty$    & $\infty$    & 2.7       & 3.7      & 1.1       \\
(Nd, La)NiO$_{2+\delta}$ buffer layer                            & 27.2       & 30.6       & 36.8       & 1.1       & 4.1       & 4.8      \\
STO (LGO) capping layer                             & 13.3       & 14.8       & 12.3      & 4.5       & 5.8       & 1.9      \\ \hline\hline
7 layers (Nd, La)NiO$_3$                            & 29.1 & 29.5 & 30.0 & 5.9 & 7.4 & 4.4 \\
7 layers of STO (LGO)                            & 15.5 & 15.0 & 15.7 & 5.6 & 6.0 & 3.9 \\
STO substrate                                 & $\infty$    & $\infty$    & $\infty$    & 3.9       & 3.7     & 2.1       \\
(Nd, La)NiO$_3$ buffer layer                            & 30.9       & 28.5      & 28.7      & 2.1       & 1.2       & 1.8     \\
STO (LGO) capping layer                             & 13.5       & 13.8       & 17.8      & 6.7      & 7.6      & 2.2      \\
\hline\hline
\end{tabular}
\end{adjustbox}
\caption{ Structural fit parameters of the reduced (upper block) and pristine samples (lower block). To reduce the number of parameters the structure of 7 out of 8 repeated $R$NO-ABO bilayers in the superlattices were coupled.}
\label{Tab_XRRpara}
\end{table*}

\section{Parameters used in the cluster calculations}
Table~\ref{Tab_Cluster} summarizes the parameters used in the cluster calculations.

\begin{table*}[h]
\begin{adjustbox}{width=1\textwidth}
\setlength{\tabcolsep}{10pt}
\renewcommand{\arraystretch}{1.2}
\centering
\begin{tabular}{l | c c c c c c c}
\hline  \hline
& $U_{dd}$ (eV) & $\Delta$ (eV) & $F^{2}_{dd}$ / $F^{4}_{dd}$ (eV)* & $\zeta_{3d}$ (eV) & $U_{pd}$ (eV) & $F^2_{pd}$ / $G^1_{pd}$ / $G^3_{pd}$ (eV) * & $\zeta_{core}$ (eV)\\
\hline
$2p^6$ $3d^8$ & 6.0 & 4.5 & 9.786 / 6.076 & 0.083 & 8.5 & 6.176 / 4.626 / 2.632 & 11.507\\
$2p^6$ $3d^9$ & 6.0 & 4.5 & 8.867 / 5.468 & 0.074 & 8.5 & 5.678 / 4.210 / 2.394 & 11.509\\
\hline\hline
\end{tabular}
\end{adjustbox}
\caption{Parameters used in the cluster calculations: $U_{dd}$ denotes the Hubbard Coulomb repulsion between electrons in the $d$ shell and $\Delta$ is the charge-transfer energy. For the Coulomb repulsion of the excited core electron $U_{pd}$ we used the values of NiO. The Slater integrals for the ground state $2p^6 3d^n$ ($F^{2}_{dd}$ / $F^{4}_{dd}$) and the state with excited core electron $2p^5 3d^{n+1}$ 8$F^2_{pd}$ / $G^1_{pd}$ / $G^3_{pd}$), as well as the spin-orbit coupling in the $3d$ ($\zeta_{3d}$) and $2p$ core-level ($\zeta_{core}$) were calculated within the Hartree-Fock approximation \cite{Haverkort2005}. Values marked with an asterisk were reduced by a factor of 0.8 from their atomic values (Refs.~\onlinecite{deGroot1990,Mizokawa1996}).}\label{Tab_Cluster}
\end{table*}

\section{Simulation of the energy dependent reflectivity}

\begin{figure*}[tb]
\center\includegraphics[width=0.98\linewidth]{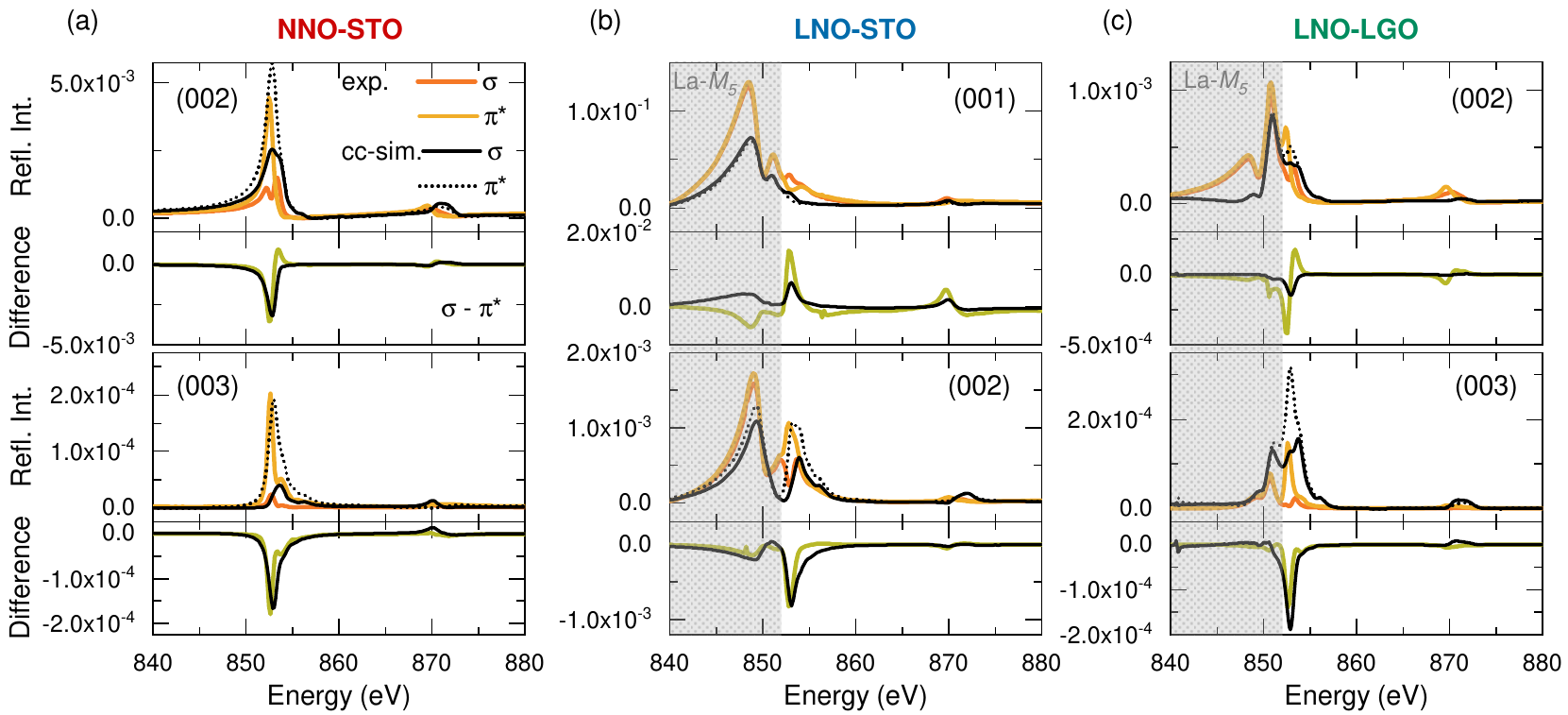}
\caption{Reflectivity constant-q$_z$ energy scans for the reduced (a) NNO-STO, (b) LNO-STO, and (c) LNO-LGO superlattices. Experimental spectra are shown by colored lines and the fits using calculated Ni$^{1+}$ and Ni$^{2+}$ from the cluster calculations are shown as black lines (cc-sim.). The other nomenclature and labeling are the same as in Fig.~\ref{Fig_XXR_NiL}.}
\label{Fig_XRR_NiL_cluster}
\end{figure*}

Figure~\ref{Fig_XRR_NiL_cluster} shows again the experimental data of Fig.~\ref{Fig_XXR_NiL} together with the calculated energy scans using the scattering tensors obtained by the cluster calculation.

%

%\bibliography{Literature}

\end{document}